\documentclass[aps,prb,
showpacs,floatfix,superscriptaddress]{revtex4}

\usepackage{amsmath,amssymb}
\usepackage[dvips]{graphics}
\usepackage{epsfig}
\usepackage{times}

\begin{document}
\title{Detection of finite frequency photo-assisted shot noise with a resonant circuit}
\author{D. Chevallier}
\affiliation{Centre de Physique Th\'eorique, UMR6207, Case 907, Luminy, 13288 Marseille Cedex 9, France}
\affiliation{Universit\'e de la M\'editerran\'ee, 13288 Marseille Cedex 9, France}
\author{T. Jonckheere}
\affiliation{Centre de Physique Th\'eorique, UMR6207, Case 907, Luminy, 13288 Marseille Cedex 9, France}
\author{E. Paladino}
\affiliation{Centre de Physique Th\'eorique, Universit\'e de la M\'editerran\'ee, Case 907, 13288 Marseille, France}
\affiliation{MATIS CNR-INFM,
and
D.M.F.C.I., Universit\'a di Catania, 95125 Catania, Italy}
\author{G. Falci}
\affiliation{Centre de Physique Th\'eorique, Universit\'e de la M\'editerran\'ee, Case 907, 13288 Marseille, France}
\affiliation{MATIS CNR-INFM,
and
D.M.F.C.I., Universit\'a di Catania, 95125 Catania, Italy}
\author{T. Martin}
\affiliation{Centre de Physique Th\'eorique, UMR6207, Case 907, Luminy, 13288 Marseille Cedex 9, France}
\affiliation{Universit\'e de la M\'editerran\'ee, 13288 Marseille Cedex 9, France}

\begin{abstract}
Photo-assisted transport through a mesoscopic conductor occurs when an oscillatory (AC) voltage is superposed to the constant (DC) bias which is imposed on this conductor. Of particular interest is the photo assisted shot noise, which has been investigated theoretically and experimentally for several types of samples. For DC biased conductors, a detection scheme for finite frequency noise using a dissipative resonant circuit, which is inductively coupled to the mesoscopic device, was developped recently. We argue that the detection of the finite frequency photo-assisted shot noise can be achieved with the same setup, despite the fact that time translational invariance is absent here. We show that a measure of the photo-assisted shot noise can be obtained through the charge correlator associated with the resonant circuit, where the latter is averaged over the AC drive frequency. We test our predictions for a point contact placed in the fractional quantum Hall effect regime, for the case of weak backscattering. The Keldysh elements of the photo-assisted noise correlator are computed. For simple Laughlin fractions,  the measured photo-assisted shot noise displays peaks at the frequency corresponding to the DC bias voltage, as well as satellite peaks separated by the AC drive frequency. 
\end{abstract}

\maketitle

\section{Introduction}

The understanding of the transport properties of nanoscale conductors at low temperatures has known tremendous successes via experiments in a wide range of systems performed for the most past in the stationary regime. Correspondingly, theoretical modelling has allowed the description of these transport processes via scattering theory approaches as well as Hamiltonian formulations, in a fruitful dialogue with experimental investigations. Transport is first characterized by the average current flowing through conductors. But further
information can be gained via the measurement and analysis of the current fluctuations\cite{blanter_buttiker,martin_houches} and more generally via the higher current moments.\cite{reulet} Early investigations of quantum transport focused almost exclusively on the low frequency regime. Few recent experiments have probed quantum system on timescales comparable with the electron correlation time, where new physical effects are expected.
The present work deals with the detection of quantum noise at such high frequencies, when both 
a DC and a AC bias is imposed between the source and the drain of the mesoscopic system.  

Indeed, high frequency measurements can mean several things. First, if only a DC bias is imposed
on the sample, a stationary current is generated and high frequencies refer to the Fourier component
of the current-current correlation function in time.\cite{yang,schoelkopf1,reydellet,chamon_freed_wen2} 
Second, high frequencies can be injected as a drive on the mesoscopic circuit,\cite{pump_kouvenhoven,bruder,lesovik_levitov,pedersen_buttiker} 
for instance when an additional AC drive is superposed to the DC bias. The later effect is called photo-assisted (PA) transport: electrons undergoing transmission from one lead to another are able to absorb/emit ``photons'' during this process.  
PA transport, and in particular PA noise has been studied theoretically and experimentally on several occasions for diffusive metals,\cite{schoelkopf1} tunnel junctions,\cite{Gabelli} normal metal/ superconductor junctions\cite{lesovik_martin_torres,schoelkopf2} as well as quantum point contacts.\cite{reydellet} The noise caracteristics then displays some structure at values of the DC bias 
which are multiple of the AC drive frequency. 

However, high requency noise detection requires special care: conventional (low) frequency  noise detection setups are often inadequate for such measurements, and one must often resort to on-chip detection schemes, or alternatively/equivalently to schemes
where a good connection to the measurement circuit is achieved through adapted impedence lines.\cite{Zakka_Bajjani} 
On chip detectors have allowed the detection of single electrons travelling through quantum dots. Such detectors and the device they probe are parts of the same quantum system and must be treated on the same footing. They bear the peculiarity that the noise which is measured is a non trivial combination of non-symmetrized noise correlators. For DC driven systems there are existing proposals to detect high frequency noise using either capacitive or inductive coupling with an on-chip circuit.\cite{aguado_kouwenhoven} 

In a recent theoretical work, a LC resonant circuit, which was coupled inductively to the mesoscopic device circuitry, was employed as a detector of both noise and higher current moments (third moment).\cite{zazunov} The description of this generic detector included its electromagnetic environement, described at a bath of harmonic oscillators with the Caldeira Legett model\cite{caldeira}. Predictions were made on the role of such a dissipative environment and on the relevance of this harmonic detector to 
capture on high frequency current moments. However, this study considered the case of a 
mesoscopic device in a stationary regime (with a DC bias only). The hypothesis of a stationary regime greatly simplies the analysis of the detection process because of time translational invariance.
The presence of an additional AC voltage drive breaks such a property. 

Given the interest in the study of time driven mesoscopic systems and in particular PA noise, 
it seems necessary to address how detection with an auxiliary circuit can be 
achieved in such situations.   
The purpose of this work is to present a high frequency detection scheme for photoassisted noise, and to illustrate it with a calculation of photoassisted noise in  a specific situation where signatures of photoassisted transport are most dramatic. For devices composed of normal metal junctions as well as superconducting/normal metal junctions, PA noise exhibits singularities at integer ratios of the DC voltage with respect to the AC frequency: the derivative of this noise exibits jumps at such locations. 
On the other hand, for a weakly pinched quantum point contact placed in the fractional quantum Hall effect regime 
(FQHE),\cite{kane_fisher,chamon_freed_wen,saleur,saminadayar,depicciotto} 
the PA noise diverges when the DC voltage -- multiplied by the filling factor -- is a multiple of the AC frequency. This much stronger singularity is a motivation for us to apply our measurement scheme to the FQHE situation. We will show that 
as in the DC case, the measured noise captures the response of the mesoscopic circuit at the resonant frequency
of the LC circuit. It exhibits a central peak at the DC voltage, which is surrounded by satellite peaks shifted 
by the AC frequency. These predictions have the potential to be tested in experiments.  

The paper is organized as follows. In Sec. II we present the model for the LC detector. We review the results for the 
charge correlator of the DC cirsuit in Sec. III and extend this discussion to the PA situation. Sec. IV is devoted
to the presentation of the QPC in the FQHE regime and its calculation of PA noise. Plots of these quantities and of the measured noise are discussed in Sec. V. We conclude in Sec. VI.


\section{Model}

The proposed setup is the same as that presented in Ref. \onlinecite{zazunov}, except for the fact that the voltage source on the mesoscopic 
device is time dependent. A lead from such device is inductively coupled to a resonant circuit
(capacitance $C$, inductance ${\cal L}$, and dissipative component $R$).
The signal which contains information about the noise of the mesoscopic circuit 
is encoded in the time correlation function of the charge on the capacitor. 

\begin{figure}[h]
	\centering
		\includegraphics[width=6cm]{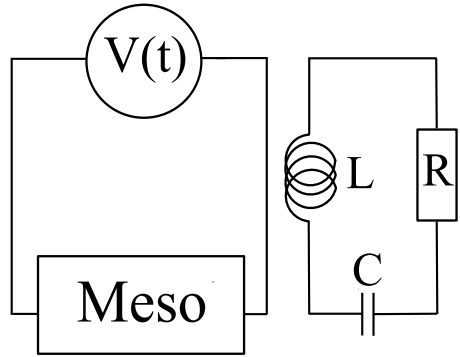}
		\caption{Mesoscopic circuit is coupled to a resonant dissipative circuit}
	\label{fig:shem}
\end{figure}

We start with the description of the detector. 
The basic Hamiltonian which describes the dissipative oscillator circuit reads:
\begin{equation}
H_{osc}=H_0+H_{LC-env}~,
\end{equation}
 where 
\begin{equation} 
H_0=H_{LC}+H_{env}
\end{equation}
is the Hamiltonian of the uncoupled
system ``$LC$ oscillator plus environment'', and $H_{LC-env}$ describes the 
coupling between the two.

For dissipative quantum systems, it is convenient to use a path integral formalism.
In the absence of dissipation and coupling to the mesoscopic device,
the Keldysh action describing the charge of the $LC$ circuit reads:
\begin{equation}
S_{LC}[q]=\frac{1}{2}\int dt dt'{\bf q}^T(t) G_0^{-1}(t-t')\sigma_z {\bf q}(t')~,
\end{equation}
where 
\begin{equation}
G_0^{-1}(t-t')={\cal L}[( i\partial_t)^2-\Omega^2] \delta(t-t')~,
\end{equation}
is the (inverse) Green function of an harmonic oscillator (${\cal L}$ is its ``mass''),
$\Omega=({\cal L}C)^{-1/2}$ is the resonant
frequency of the circuit, ${\bf q}^T=(q^+,q^-)$ is a two component vector which contains
the oscillator coordinate on the forward/backward contour, and $\sigma_z$ is a Pauli matrix in Keldysh space.
Dissipative effects are treated within the Caldeira-Leggett model,
where the environment is modeled by a set of harmonic oscillators (bath)
with frequencies $\{ \omega_n \}$;
the coordinate $q$ is coupled linearly to the bath oscillators: 
\begin{equation}
H_{LC-env} = q \sum_n \lambda_n x_n~,
\end{equation}
with the coupling constants $\lambda_n$.

The partition function of the $LC$ oscillator plus bath,
$Z=\int\mathcal{D}q \mathcal{D}x e^{ i S[q,x]}$, has an action:
\begin{equation}
S = S_{LC}+ \frac{1}{2}\sum_n{\bf x}^T_n\circ D_n^{-1}\circ \sigma_z {\bf x}_n  -
{\bf q}^T\circ\sigma_z \sum_n\lambda_n {\bf x}_n~,
\end{equation}
where $D^{-1}_n(t)=M_n[( i\partial_t)^2-\omega_n^2]\, \delta(t)$ and
the symbol $\circ$ stands for convolution in time.
The bath degrees of freedom can be integrated out in a standard manner \cite{grabert_report}.
As a result, the Green function $G$ of the $LC$ circuit becomes dressed by its electronic environment,
\begin{equation}
G^{-1} = G_0^{-1} - \Sigma~,
\end{equation}
with a self-energy $\Sigma(t)=\sigma_z \sum_n \lambda_n^2 D_n(t) \sigma_z$. In the remainer of this paper, when we mention the $LC$ circuit, it will also imply the presence of its surrounding electromagnetic environment.

Next, we introduce the inductive
coupling between the mesoscopic device and the $LC$ circuit,
\begin{equation}
H_{int}=\alpha q \dot{I}~,
\end{equation} 
where $\dot{I}$ is the time derivative
current operator \cite{lesovik_loosen}. This interaction is interpreted
here as an external potential acting on the oscillator circuit.
To calculate correlation functions of the $LC$ circuit coordinate $q$,
we introduce the generating functional,
\begin{equation}
\mathcal{Z}[\textrm{\boldmath $\eta$},{\bf I}]=\int \mathcal{D}{\bf q} \exp i
\Big[\frac{1}{2} \, {\bf q}^T\circ G^{-1} \circ {\bf q} -{\bf q}^T\sigma_z\circ(\alpha \dot{\bf I}+
\textrm{\boldmath $\eta$})\Big]~,
\end{equation}
where $\textrm{\boldmath $\eta$}^T=(\eta^+,\eta^-)$ is a two-component auxiliary field.
Performing integration over the LC oscillator variables ${\bf q}$ results in 
$\mathcal{Z}[\textrm{\boldmath $\eta$},{\bf I}] = e^{i S_{eff}[\textrm{\boldmath $\eta$},{\bf I}]}$
with an effective action (restoring integrals):
\begin{eqnarray}
S_{eff}[\textrm{\boldmath $\eta$}, {\bf I}]&=&-\frac{ i}{2}\int dt \int dt' (\textrm{\boldmath $\eta$}(t)+\alpha\dot{\bf I}(t))^T\sigma_z
\check{G}(t-t')\nonumber\\
&&~~~~~~\times \sigma_z (\textrm{\boldmath $\eta$}(t')+\alpha\dot{\bf I}(t'))\Big]~.
\label{action_integrated}
\end{eqnarray}

\section{Charge correlator}
\label{charge_correlator}

By taking double derivatives of the Kelysh partition function with 
respect to the components of the spinor $\eta$, 
the charge correlator is obtained:
\begin{eqnarray}
\langle q^\beta(t)q^{\beta'}(t')\rangle \equiv  \mathcal{Z}^{-1}_\eta[I]
\frac{\partial^2\mathcal{Z}_\eta[I]}{\partial \eta(t^{\beta})\partial \eta(t'^{\beta'})}_{\eta=0}~,
\end{eqnarray}
where $\beta,\beta'\equiv \pm 1$ are indices specifying the upper/lower branch of the Keldysh contour.  
To leading order in the coupling constant $\alpha$ between the 
mesoscopic circuit and the detector\cite{zazunov}
this can be expressed in terms of the current derivative correlator:
\begin{equation}
K^{\beta_1\beta_2}(\tau_1,\tau_2)=\left\langle T_K \dot{I}(\tau_{1})^{\beta_1}\dot{I}(\tau_{2})^{\beta_2}\right\rangle_{meso}
~,
\label{current_derivative}\end{equation}
where the average $\left\langle ...\right\rangle_{meso}$ represents a non equilibrium average
containing information on the occupation of the reservoirs connected to the sample 
and on its scattering properties. 
The charge correlator consists then of a Keldysh matrix:
\begin{align}
\left\langle T_{K} q^{\beta}(t) q^{\beta'}(t') \right\rangle =\alpha^2 \int d\tau_1 d\tau_2 \sum_{\beta_1\beta_2} G^{\beta\beta_2}(t-\tau_2)\sigma^{\beta_2\beta_2}_z K^{\beta_2\beta_1}(\tau_2,\tau_1)\sigma^{\beta_1\beta_1}_z G^{\beta_1\beta'}(\tau_1-t')~,
\label{keldysh_charge_correlator}
\end{align}
where the integrand contains the Green function $G^{\beta\beta'}(t)$ of the LC circuit only.
While this Green function is a function of a single time argument because of time 
translational invariance, the current derivative correlator $K^{\beta_1\beta_2}(\tau_1,\tau_2)$ is 
not a function of the difference  $\tau_1-\tau_2$ if the bias voltage is time dependent. 

\subsection{DC Voltage only}

We recall the results obtained previously for the detection of finite frequency noise 
in the presence of time translational invariance. The initial proposal 
of Ref. \onlinecite{lesovik_loosen} for a dissipationless LC circuit was to operate
repeated time measurements on the charge $q$.
This allows to construct an histogram for zero voltage, yielding
the zero bias peak position, its width, skewness,... In the presence of bias,
this histogram is shifted, and acquires a new width, skewness,... Information
about all current moments at high frequencies is encoded in such histograms.
Here, however, we only focus on the detection of noise.
In Ref. \onlinecite{zazunov}, the inclusion of dissipation due to 
the electromagnetic environment was shown to be essential to obtain a finite 
result for the measuring process. 
There, expressions for the off diagonal Keldysh 
component of the charge correlator
$\langle T_K q^-(t)q^{+}(t')\rangle=\langle q(t-t')q(0)\rangle$ were derived
with the help of Eq. (\ref{keldysh_charge_correlator}). Note that in this situation, the 
current derivative correlator of Eq. (\ref{current_derivative}) is a function
of the difference $\tau_1-\tau_2$, and the charge correlator is a convolution product, 
which explains its dependence on $t-t'$ only. 

Going to the rotated Keldysh basis (see Appendix \ref{rakbasis}) allows to rewrite the charge fluctuations at equal time ($t=t'$) as:
\begin{eqnarray}
\delta\langle q^2 \rangle &=& \alpha^2 \int \frac{d\omega}{2\pi} \, G^R(\omega) \{G^K(\omega)K^{+-}(\omega)-(G^R(\omega)-G^A(\omega))K^{-+}(\omega)\}~,
\label{general_second}
\end{eqnarray}
with the three Green function components given by:
\begin{equation}
G^{R/A}(\omega) = [{\cal L}(\omega^2-\Omega^2)\pm i \,\textrm{sgn}(\omega)J(|\omega|)]^{-1}~,
\end{equation}
and
\begin{equation}
G^K=(2N(\omega)+1)(G^R(\omega)-G^A(\omega))~,
\end{equation}
where $N(\omega)$ is the Bose occupation number of the oscillator and the bath spectral function is defined as:
\begin{equation}
J(\omega)=\pi\sum_n\lambda_n^2/(2M_n\omega_n)\delta(\omega-\omega_n)~.
\end{equation}
This spectral function is at the origin of the broadening 
for the $LC$ circuit Green function.

The time derivative correlators $K^{-+,+-}$ are related to
the Fourier transform of the current current correlation functions as
$K^{+-}(\omega)=\omega^2S^{+-}(\omega)$ and $K^{-+}(\omega)=\omega^2S^{-+}(\omega)$,
with 
\begin{equation}
S^{+-}(\omega)= \int dt \langle I(0) I(t)\rangle e^{i\omega t}~,
\end{equation}
 and
$S^{-+}(\omega)=S^{+-}(-\omega)$ corresponding to the response function for
emission/absorption of radiation from/to the mesoscopic circuit \cite{lesovik_loosen,deblock_science}.
With these definitions, the final result for the measurable excess noise reads:
\begin{eqnarray}
\delta\langle q^2 \rangle&=&
  2\alpha^2 \int_{0}^{\infty} \frac{d\omega}{2\pi}
 \omega^2
 [\chi''(\omega)]^2
\nonumber\\
&&~~\times
  \big(S^{+-}(\omega)+N(\omega)(S^{+-}(\omega)-S^{-+}(\omega))\big)~,
\label{ohmic_second}
\end{eqnarray}
where $\chi''(\omega)= J(|\omega|)/[{\cal L}^2(\omega^2-\Omega^2)^2 + J^2(|\omega|)$
is the susceptibility of Ref. \onlinecite{caldeira}, here generalized to arbitrary
$J(|\omega|)$.
Eq. (\ref{ohmic_second}) constitutes a mesoscopic analog of the radiation line width
calculation\cite{larkin}: a dissipative $LC$ circuit cannot yield any divergences
in the measurable noise. Dissipation is essential in the measurement process.

Eq. (\ref{ohmic_second}) indicates that for an infinitesimal line width,
the integrand can be computed at the resonant frequency $\Omega$, and the
measured noise takes the form of Ref. \onlinecite{lesovik_loosen}:
\begin{equation}
\left\langle q^{2}\right\rangle=\frac{\alpha^{2}}{\gamma {\cal L}^{2}}\left\{S^{+-}(\Omega)+N_{\Omega}(S^{+-}(\Omega)-S^{-+}(\Omega))\right\}~,
\end{equation}
where the prefactor $\gamma$ is defined assuming a strict Ohmic or Markovian 
damping ($J(\omega)= {\cal L} \gamma \omega$), which corresponds to 
a memoryless bath which is consistent with the adiabatic switching assumption, as discussed in Ref. \onlinecite{zazunov}. 

As an alternative to the measurement of the width of the charge distribution, 
one can imagine that the capacitor itself is coupled to a measuring device (a single electron tunneling device) which directly detects the Fourier transform of the charge correlator.\cite{per_delsing} 
Given the fact that the charge correlator matrix of Eq. (\ref{keldysh_charge_correlator})
is a convolution product in this stationary situation, its Fourier transform take the 
simple form of a product of matrices:
\begin{equation}
\label{product_fourier}
\int dt  \langle T_K q^\beta(t)q^{\beta'}(0)\rangle
e^{i\omega t}=\alpha^2\left[\tilde{G}(\omega)\sigma_{z}\tilde{K}(\omega)\sigma_{z}\tilde{G}(\omega)
\right]^{\beta\beta'}~,
\end{equation}
where $\tilde{G}(\omega)$ and $\tilde{K}(\omega)$ are respectively the matrix version 
of the LC Green's function and of the current derivative correlator. 
Naturally this will have substantial contributions when both $K$ and $G$ overlap significantly.
This constitutes a rather compact way for describing the detection process in the case of 
a constant bias voltage.   

\subsection{AC drive and temporal invariance}

We now turn to the main point of this section, which is to address how to deal with the presence 
of an AC voltage superposed to the DC one. The total bias potential $V(t)$ which is 
applied to the mesoscopic device is thus a periodic function of time 
with period $\tau=2\pi/\omega_{AC}$.
We start by defining a correlator $k(T,t'')$ 
from the current derivative correlator of Eq. (\ref{current_derivative}):
\begin{equation}
K(t,t')\equiv k(\frac{t+t'}{2},t-t')~.
\end{equation}
Defining $T=(t+t')/2$, $t''=t-t'$, the charge correlator of 
Eq. (\ref{keldysh_charge_correlator}) is rewritten as: 
\begin{equation}
\left\langle T_K q^{\beta}(t) q^{\beta'}(t') \right\rangle =\alpha^2 \int dt_1 dt_2 \sum_{\beta_1\beta_2} G^{\beta\beta_2}(t'' -t_2)\sigma^{\beta_2\beta_2}_z k^{\beta_2\beta_1}(T+t_0,t_2-t_1)\sigma^{\beta_1\beta_1}_z G^{\beta_1\beta'}(t_1)~.
\end{equation}
where $t_0=(t_2+t_1)/2-t''/2$.
Next, we define the average of the charge correlator over the period of the AC drive\cite{pedersen_buttiker}
as follows: 
\begin{equation}\label{upo}
\frac{1}{\tau}\int^{\tau}_{0}dT\left\langle q^{\beta}(t) q^{\beta'}(t') \right\rangle=\alpha^2 \int dt_1 dt_2 \sum_{\beta_1\beta_2}  
G^{\beta\beta_2}(t'' -t_2)\sigma^{\beta_2\beta_2}_z \int^{\tau}_{0}\frac{dT}{\tau} k^{\beta_2\beta_1}(T+t_0,t_2-t_1)\sigma^{\beta_1\beta_1}_zG^{\beta_1\beta'}(t_1)~.
\end{equation}
Note that the last integral over the variable $T$ is essentially a period average of the correlator 
$k(T,t'')$ with the variable $T$ shifted by $t_0$.
In the presence of an AC drive, this period average does not depend on the shift $t_0$, 
because as a function of the variable $T$, $k(T,t'')$ contains only harmonics of the drive 
frequency  $\omega_{AC}$. This has been noticed in earlier works.
\cite{lesovik_martin_torres,crepieux_photoassisted,guigou} For our purposes, it means 
that we can safely replace $t_0$ by $0$.  As a result, the period averaged charge correlator
takes the form of a convolution product as was the case for the constant DC bias, and it therefore
depends only on the time difference $t-t'$:
\begin{eqnarray}
\label{Bruit}
\hat{Q}^{\beta\beta'}(t-t')&\equiv&\frac{1}{\tau}\int^{\tau}_{0}dT \left\langle q^{\beta}(t) q^{\beta'}(t') \right\rangle
\nonumber\\
&=&
\alpha^2 \int dt_1 dt_2 \sum_{\beta_1\beta_2}  
G^{\beta\beta_2}(t'' -t_2)\sigma^{\beta_2\beta_2}_z {\cal{K}}^{\beta_2\beta_1}(t_2-t_1)
\sigma^{\beta_1\beta_1}_z G^{\beta_1\beta'}(t_1)~,
\end{eqnarray}
where we defined the period averaged correlator:
\begin{equation}\label{period_averaged_correlator}
{\cal{K}}^{\beta_2\beta_1}(t)\equiv \frac{1}{\tau}\int^{\tau}_0 dT k^{\beta_2\beta_1}(T,t)~.
\end{equation}
Finally, the averaged charge correlator can be expressed in terms of the Fourier transfrom of both the 
LC circuit Green's function and the period averaged current correlator:
\begin{equation}
\label{average_charge_correlator_final}
\hat{Q}^{\beta\beta'}(t'')=\alpha^2\sum_{\beta_1\beta_2} \int\frac{d\omega}{2\pi}e^{-i\omega t''}G^{\beta\beta_2}(\omega)\sigma^{\beta_2\beta_2}_z {\cal{K}}^{\beta_2\beta_1}(\omega)\sigma^{\beta_1\beta_1}_zG^{\beta_1\beta'}(\omega)~.
\end{equation}
This result is the exact analog of the DC formula Eq. \ref{product_fourier}, extended to and AC drive.
In addition, at $t''=0$, this expression has the 
same form as the result of Ref. \onlinecite{zazunov}.
We have therefore identified which quantity (${\cal K}$) characterizes 
the influence of the mesoscopic circuit on the response of the 
LC circuit. Therefore, the protocol for measuring photoassisted shot noise
is the same as in the DC case provided one averages the response over 
the frequency of the drive. This averaging procedure restores the 
temporal invariance of the charge correlator. In the following sections, we will
compute the current derivative correlators and their period average 
for a specific system: a QPC placed in the conditions of the FQHE 
where the elementary transport process is the Poissonian transfer of 
Laughlin quasiparticles.  

\section{Non symmetrized photo-assisted noise in the FQHE}

The calculation of the symmetrized photoassisted noise has been carried 
out in Ref. \onlinecite{crepieux_photoassisted}. Here we use the same basic model
and generalize the calculations of the noise correlator to the full Keldysh
matrix elements of this correlator. Next, we extract from these the noise 
derivative correlators which are relevant for the measurement process. 

\subsection{Model for quasiparticle backscattering}

\begin{figure}[ht]
	\centering
		\includegraphics[width=5cm]{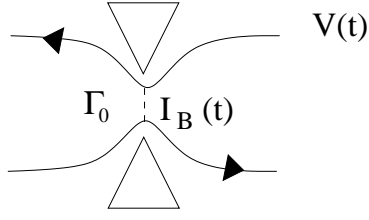}
	\caption{Quantum point contact}
	\label{fig:QPC}
\end{figure}

We use the Tomonaga-Luttinger formalism to describe the 
right and left moving chiral excitations.
In the absence of tunneling between the two edges, 
the Hamiltonian reads:
\begin{equation}
H_{0}=(\frac{\nu_{F}\hbar}{4\pi})\sum_{r}\int ds(\partial_{s}\phi_r)^2~, 
\end{equation}
with $r=+,-$ for right and left movers. 
Here, we focus solely on the weak backscattering regime because 
it is already known that the PA shot noise exhibits some 
strong singularities. 
The backscattering of quasiparticles is described by the Hamiltonian:
\begin{eqnarray}
H_B(t)=\sum_{\varepsilon} A^{(\varepsilon)}(t)[\Psi_+^\dag(t)\Psi_-(t)]^{(\varepsilon)}~,
\end{eqnarray}
where $A^{(\varepsilon)}(t)$ is a tunneling amplitude which depends
on the applied voltage via the Peierls substitution.
Here the notation $\epsilon=\pm$ leaves an operator unchanged for ($\epsilon=+$) 
or specifies its Hermitian conjugate ($\epsilon=-$). 
$\Psi_{r}$ is the quasiparticle operator which is expressed in terms of the 
bosonic chiral field $\phi_{r}$:
\begin{eqnarray}
\Psi_r(t)=\frac{1}{\sqrt{2\pi a}}e^{i\sqrt{\nu}\phi_r(t)}~,
\end{eqnarray}
where $a$ is a short distance cutoff and $\nu$ is the filling factor ($\nu^{-1}$ is an odd integer to describe Laughlin fractions). 
Choosing a time dependent voltage in the form $V(t)=V_0+V_1\mathrm{cos}(\omega_{AC} t)$ results in a tunneling amplitude:
\begin{eqnarray}
A^{(\varepsilon)}(t)=\Gamma_0e^{i\varepsilon\omega_0t}\mathrm{exp}(i\varepsilon \frac{ e^*V_1}{\hbar \omega_{AC}}\mathrm{sin}(\omega_{AC} t))~,
\end{eqnarray}
where $e^{*}=\nu e$ and $\Gamma_0$ is the bare tunneling amplitude. 
The backscattering current is deduced from the backscattering Hamiltonian:
\begin{eqnarray}
I_B(t)=\frac{ie^*}{\hbar}\sum_{\varepsilon}\varepsilon A^{(\varepsilon)}(t)[\Psi_+^\dag(t)\Psi_-(t)]^{(\varepsilon)}
~.\end{eqnarray}

\subsection{Non-symmetrized noise}

The general expression for the Keldysh components of the noise correlator in the Heisenberg representation is:
\begin{eqnarray}
S^{\beta\beta'}(t,t')=\langle I^{\beta}_B(t)I^{\beta'}_B(t')\rangle -\langle I_B(t^{\beta})\rangle\langle I_B(t'^{\beta'})\rangle~.
\end{eqnarray}
Since we are interested in Poissonian regime only, the product of current averages
can be dropped out because it contributes to higher order in the backscattering Hamiltonian.\cite{martin_houches} 
Moreover, in this second order calculation in the tunneling amplitude $\Gamma_0$, 
there is no difference between the Heisenberg and interaction picture. 
The noise then reads:  
\begin{eqnarray}
S^{\beta\beta'}(t,t')
=-(e^*)^2\sum_{\varepsilon\varepsilon'}\varepsilon\varepsilon'
A^{(\varepsilon)}(t)A^{(\varepsilon')}(t')
\langle T_K\{ [\Psi_+^\dag(t^\beta)\Psi_-(t^\beta)]^{(\varepsilon)} [\Psi_+^\dag(t'^{\beta'})\Psi_-(t'^{\beta'})]^{(\varepsilon')}\}\rangle
~.\end{eqnarray}
This correlator is different from zero only when $\varepsilon'=-\varepsilon$ because of 
quasiparticle conservation. 
Replacing the quasiparticle correlators by their bosonized expression, the noise is 
then written in terms of a product over averages of bosonic fields:   
\begin{eqnarray}
S^{\beta\beta'}(t,t')
=\frac{(e^*)^{2}}{4\pi^2a^2}
\sum_{\varepsilon} \varepsilon 
A^{(\varepsilon)}(t)A^{(-\varepsilon)}(t')
\langle T_K 
e^{-i\varepsilon\sqrt{\nu}\phi_+(t^\beta)}
e^{i\varepsilon\sqrt{\nu}\phi_+(t'^{\beta'})}
\rangle
\langle T_K 
e^{i\varepsilon\sqrt{\nu}\phi_-(t^\beta)}
e^{-i\varepsilon\sqrt{\nu}\phi_-(t'^{\beta'})}
\rangle
~.\end{eqnarray}
The final result for the real time noise correlator is then:
\begin{eqnarray}
S^{\beta\beta'}(t,t')
=\frac{(e^*)^2}{4\pi^2a^2}
e^{2\nu G^{\beta\beta'}(t-t')}\left(A(t)A^{*}(t')+A^*(t)A(t')\right)
~,\end{eqnarray}
where we introduced the chiral green function of the bosonic fields:
\begin{eqnarray}
G^{\beta\beta'}(t,t')=\langle T_K\{\phi_r(t^\beta)\phi_r({t'}^{\beta'})\}\rangle
-\frac{1}{2}\langle T_K\{\phi_r(t^\beta)^2\}\rangle
-\frac{1}{2}\langle T_K\{\phi_r({t'}^{\beta'})^2\}\rangle
~.\end{eqnarray}
The double Fourier transform of this quantity, which will allow to relate it to the noise 
correlator, reads:
\begin{eqnarray}
S^{\beta\beta'}(\Omega_1,\Omega_2)= \int\int dtdt' e^{i(\Omega_1t+\Omega_2 t')}S(t,t')~.  
\end{eqnarray}
We now specify the periodic voltage modulation, which allows to write the tunneling 
amplitude in terms of a series of Bessel functions $J_n$:
\begin{eqnarray}
A(t)=\Gamma_0\sum_{n=-\infty}^{+\infty}e^{i(\omega_0+n\omega_{AC})t}
J_n\left(\frac{e^*V_1}{\hbar\omega_{AC}}\right)~,
\end{eqnarray}
which gives the Fourier transform of non-symmetrised noise:
\begin{eqnarray}
S^{\beta\beta'}(\Omega_1,\Omega_2)&=&\frac{(e^*)^2\Gamma_0^2}{2\pi^2a^2}\sum_{n=-\infty}^{+\infty}
\sum_{m=-\infty}^{+\infty}J_n\left(\frac{e^*V_1}{\omega_{AC}}\right)
J_m\left(\frac{e^*V_1}{\omega_{AC}}\right)\nonumber\\
&&\times\int\int dt dt'e^{i(\Omega_1 t+\Omega_2 t')}
e^{2\nu G^{\beta\beta'}(t,t')}\mathrm{cos}(\omega_0(t-t')+\omega_{AC}(nt-mt'))~.
\end{eqnarray}
Next, it is convenient to perform a 
change of variable $\tau=t-t'$ and $\tau'=t+t'$:
\begin{eqnarray}\label{variable_change}
S^{\beta\beta'}(\Omega_1,\Omega_2)&=&2\frac{(e^*)^2\Gamma_0^2}{2\pi^2a^2}\sum_{n=-\infty}^{+\infty}
\sum_{m=-\infty}^{+\infty}J_n\left(\frac{e^*V_1}{\omega_{AC}}\right)
J_m\left(\frac{e^*V_1}{\omega_{AC}}\right)\nonumber\\
&&\times\int\int d\tau d\tau'e^{i(\Omega_1-\Omega_2)\tau/2}e^{i(\Omega_1+\Omega_2)\tau'/2}
e^{2\nu G^{\beta\beta'}(\tau)} \mathrm{cos}\left(\left(\omega_0+\frac{n+m}{2}\omega_{AC}\right)\tau+\frac{n-m}{2}\omega_{AC}\tau'\right)~.\notag
\end{eqnarray}
Using standard trigonometric identities, one can write this expression as a 
product of separate integrals over $\tau$ and $\tau'$. Integrals over $\tau$ contain
the (zero temperature) Green's function of the chiral fields and can be expressed in terms of 
Gamma function. 
The result has the form:
\begin{eqnarray}
S^{\beta\beta'}(\Omega_1,\Omega_2)&=&2\frac{(e^*)^2\Gamma_0^2}{2\pi^2a^2}\sum_{n=-\infty}^{+\infty}
\sum_{m=-\infty}^{+\infty}J_n\left(\frac{e^*V_1}{\omega_{AC}}\right)
J_m\left(\frac{e^*V_1}{\omega_{AC}}\right)\nonumber\\
&~~&~~\left[I_{1}(\Omega_1+\Omega_2,\omega)
I_{2}^{\beta\beta'}(\Omega_1-\Omega_2,\omega_0,\omega)
-
I_{3}(\Omega_1+\Omega_2,\omega)
I_{4}^{\beta\beta'}(\Omega_1-\Omega_2,\omega_0,\omega)
\right]
~.\end{eqnarray}
The integrals $I_1$, $I_{2}^{\beta\beta'}$, $I_3$, $I_{4}^{\beta\beta'}$ are defined and computed in the Appendix. 
The final result for the 4 Keldysh matrix elements 
of the noise correlator is:

\begin{align}
S^{\beta-\beta}(\Omega_1,\Omega_2)&=2\frac{(e^*)^2\Gamma_0^2}{4\pi^2a^2}\sum^{+\infty}_{n=-\infty}
\sum^{+\infty}_{m=-\infty}J_n\left(\frac{e^*V_1}{\omega_{AC}}\right)
J_m\left(\frac{e^*V_1}{\omega_{AC}}\right)\frac{\pi}{\Gamma(2\nu)}(\frac{a}{\nu_{F}})^{2\nu}\times\notag\\
&\left[\left(1-\beta \mathrm{sgn}\left(\Omega_1+\omega_0+n\omega_{AC}\right)\right)\left|\Omega_1+\omega_0+n\omega_{AC}\right|^{2\nu-1}\delta(\Omega_1+\Omega_2+(n-m)\omega_{AC})\right.\notag\\
&+\left.\left(1-\beta \mathrm{sgn}(\Omega_1-\omega_0-n\omega_{AC})\right)\left|\Omega_1-\omega_0-n\omega_{AC}\right|^{2\nu-1}\delta(\Omega_1+\Omega_2-(n-m)\omega_{AC})\right]
~,\end{align}

\begin{align}
S^{\beta\beta}(\Omega_1,\Omega_2)&=2\frac{(e^*)^2\Gamma_0^2}{4\pi^2a^2}\sum^{+\infty}_{n=-\infty}
\sum^{+\infty}_{m=-\infty}J_n\left(\frac{e^*V_1}{\omega_{AC}}\right)
J_m\left(\frac{e^*V_1}{\omega_{AC}}\right)\frac{\pi}{\Gamma(2\nu)}(\frac{a}{\nu_{F}})^{2\nu}\frac{ e^{-\beta i\pi\nu}}{\mathrm{cos}(\pi\nu)}\times\notag\\
&\times\left[\left|\Omega_1+\omega_0+n\omega_{AC}\right|^{2\nu-1}\delta(\Omega_1+\Omega_2+(n-m)\omega_{AC})+\left|\Omega_1-\omega_0-n\omega_{AC}\right|^{2\nu-1}\delta(\Omega_1+\Omega_2-(n-m)\omega_{AC})\right]
~.\end{align}

We recognize that since we are dealing with simple Laughlin fractions of the FQHE, $\nu$ is the inverse of an odd integer
and all Keldysh component exhibit power law singularities when the quantity
$\Omega_2\pm(\omega_0+n\omega_{AC})$ vanishes. As a check, it is possible to recover
from these components the previous result for the symmetrized noise:\cite{crepieux_photoassisted} 
\begin{eqnarray}
S_{sym}(\Omega_1,\Omega_2)&=&\frac{1}{2}(S^{+-}(\Omega_1,\Omega_2)+S^{-+}(\Omega_1,\Omega_2))
\nonumber\\
&=&2\frac{(e^*)^2\Gamma_0^2}{4\pi^2a^2}\sum^{+\infty}_{n=-\infty}
\sum^{+\infty}_{m=-\infty}J_n\left(\frac{e^*V_1}{\omega_{AC}}\right)
J_m\left(\frac{e^*V_1}{\omega_{AC}}\right)\frac{\pi}{\Gamma(2\nu)}(\frac{a}{\nu_{F}})^{2\nu}\nonumber\\
&~~&\times\left[\left|\Omega_1+\omega_0+n\omega_{AC}\right|^{2\nu-1}\delta(\Omega_1+\Omega_2+(n-m)\omega_{AC})\right.\nonumber\\
&~~&~~+\left.\left|\Omega_1-\omega_0-n\omega_{AC}\right|^{2\nu-1}\delta(\Omega_1+\Omega_2-(n-m)\omega_{AC})\right]
~.\end{eqnarray}
It is also useful to know that the standard property 
of Keldysh Green's functions:
\begin{equation}
S^{++}(\Omega_1,\Omega_2)+S^{--}(\Omega_1,\Omega_2)=S^{-+}(\Omega_1,\Omega_2)+S^{+-}(\Omega_1,\Omega_2)
\end{equation}
applies as it should for the double Fourier transform expressions. 

\subsection{Current derivative correlators}

The relation between the Fourier components of the noise correlator computed in the previous section and the current derivative correlator introduced in Sec. \ref{charge_correlator} reads:
\begin{equation}
K^{\beta\beta'}(\Omega_1,\Omega_2)=-\Omega_1\Omega_2 S^{\beta\beta'}(\Omega_1,\Omega_2)
~.\end{equation}
Yet, we need to relate the noise correlator $S^{\beta\beta'}(\Omega_1,\Omega_2)$ to the correlator
$k(T,\omega)$ and ultimately, to its time average ${\cal{K}}^{\beta_2\beta_1}(\omega)$.
This is achieved using the relation:
\begin{equation}
k^{\beta\beta'}(T,\omega)=\int\frac{d\omega_1}{2\pi}e^{-i\omega_1 T} K^{\beta\beta'}(\frac{\omega_1}{2}+\omega,\frac{\omega_1}{2}-\omega)
~.\end{equation} 
So the final result for the four averaged noise derivative correlators reads:
\begin{align}\label{dericorr}
{\cal{K}}^{\beta-\beta}(\omega)&=\frac{1}{\tau}\int^{\tau}_{0}dT k^{+-}(T,\omega)=\frac{(e^*)^2\Gamma_0^2}{4\pi^2a^2}\sum^{+\infty}_{n=-\infty}
J^2_n\left(\frac{e^*V_1}{\omega_{AC}}\right)
\frac{1}{\Gamma(2\nu)}(\frac{a}{\nu_{F}})^{2\nu}\omega^2\notag\\
&\times\left[\left(1-\beta\mathrm{sgn}\left(\omega+\omega_0+n\omega_{AC}\right)\right)\left|\omega+\omega_0+n\omega_{AC}\right|^{2\nu-1}
+\left(1-\beta\mathrm{sgn}\left(\omega-\omega_0-n\omega_{AC}\right)\right)\left|\omega-\omega_0-n\omega_{AC}\right|^{2\nu-1}\right]
\end{align}

\begin{align}\label{dericorre2}
{\cal{K}}^{\beta\beta}(\omega)&=\frac{1}{\tau}\int^{\tau}_{0}dT k^{++}(T,\omega)=\frac{(e^*)^2\Gamma_0^2}{4\pi^2a^2}\sum^{+\infty}_{n=-\infty}
J^2_n\left(\frac{e^*V_1}{\omega_{AC}}\right)
\frac{1}{\Gamma(2\nu)}(\frac{a}{\nu_{F}})^{2\nu}\frac{ e^{-\beta i\pi\nu}}{\mathrm{cos}(\pi\nu)}\omega^2
\notag\\
&\times\left[\left|\omega+\omega_0+n\omega_{AC}\right|^{2\nu-1}+\left|\omega-\omega_0-n\omega_{AC}\right|^{2\nu-1}\right]~.
\end{align}

To be complete, we can compute all the Keldysh element in the rotated basis. This is performed
in Appendix \ref{rakbasis}. While the advanced and retarded contribution do not bear information on the 
non equilibrium nature of the transport processes taking place in the mesoscopic devices and therefore in the detector, the Keldysh component:
\begin{equation}
\label{average_charge_correlator_final_rotated}
\tilde{Q}^K=\alpha^2\int\frac{d\omega}{2\pi}\left[\tilde{G}^R(\omega)\tilde{{\cal{K}}}^R(\omega)\tilde{G}^K(\omega)+\tilde{G}^R(\omega)\tilde{{\cal{K}}}^K(\omega)\tilde{G}^A(\omega)+\tilde{G}^K(\omega)\tilde{{\cal{K}}}^A(\omega)\tilde{G}^A(\omega)\right]
~,\end{equation}
summarizes such an information in a compact way. 
We recall that as an alternative to the measurement of the equal time charge correlator, 
Eqs. (\ref{average_charge_correlator_final}) and (\ref{average_charge_correlator_final_rotated}) are also 
likely to be measured directly (resolved in frequency) by an SET device.\cite{per_delsing}

This completes the calculation of the current derivative correlators. 
In the following section we continue with the same analysis as with the DC case\cite{zazunov}. 
That is we use the contour ordered elements of the charge correlator, in particular the 
$-+$ component evaluated at equal time: in Sec.V we insert the expressions for Eqs. (\ref{dericorr}) - (\ref{dericorre2}) and discuss the results.
 
\section{Results}

We now discuss the formulas obtained in Sec. IV. In all of the results below,
we have checked that when the AC drive frequency is set to $0$, we recover
the DC results for the finite frequency noise at $\nu=1$ [\onlinecite{martin_houches}] 
and $\nu=1/3$ [\onlinecite{safi_bena}]. For the QPC we focus on the voltage dominated regime where the temperature is
taken to zero in the current correlator, but it nevertheless enters the detector response.

\subsection{Excess noise in the quantum Hall effect}
\label{excess_subsection}

\begin{figure}[ht]
	\centering
		\includegraphics[width=18cm]{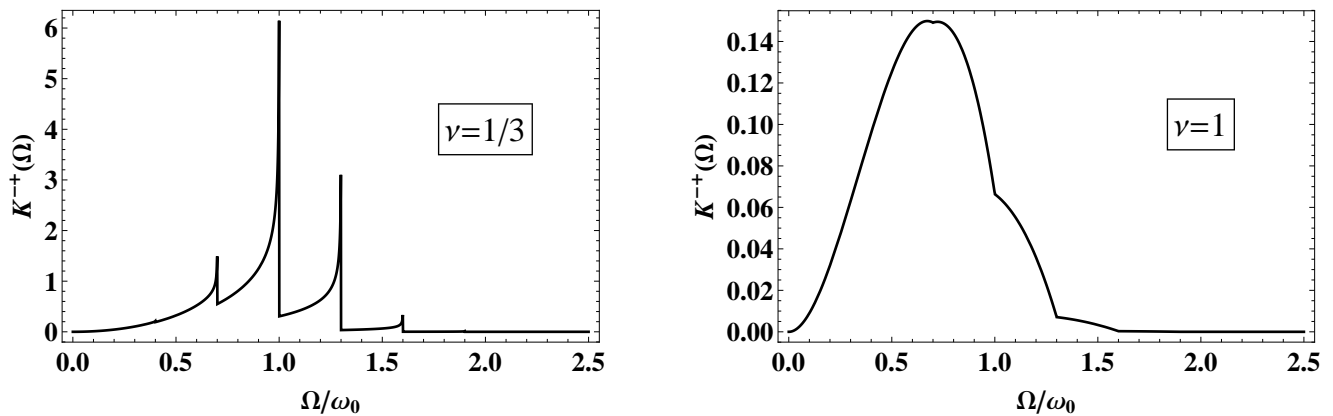}
	\caption{Current derivative correlator for a QPC in the fractional (left) and integer (right)
	quantum Hall effect ($\omega_0=3\, \omega_{AC}$,  $\omega_1=\omega_{AC}$ and ${\cal{K}}^{-+}(\Omega)$ is normalized by $e^* I_B \omega^2_0$)}
	\label{fig:Bruitderive}
\end{figure}

We start with a discussion of the results for the non symmetrized excess PA noise. 
We show in Fig. \ref{fig:Bruitderive} the curves for the current derivative correlator 
${\cal{K}}^{-+}(\Omega)$ (see eq. \ref{dericorr}), which is the quantity 
which enters the expression of the measured noise (the charge correlator). 
This is displayed 
for two different values of the filling factor $\nu$. We choose for our main interest the 
Laughlin fraction $\nu=1/3$, which is in principle the easiest attainable Laughlin 
fraction of the FQHE in experiments, 
and $\nu=1$, the integer quantum Hall effect case, which here also corresponds to the noise caracteristics
of a single channel normal tunneling junction.

Here we have chosen the DC voltage so that the central frequency $\omega_0\equiv \nu e V_0/\hbar$ 
is larger than the drive frequency $\omega_{AC}$, and the amplitude 
of the AC voltage ($\omega_1=\nu e V_1/\hbar$) is such that $\omega_1/\omega_{AC}=1$

For $\nu=1/3$ we find divergences for ${\cal{K}}^{-+}(\Omega)$ located at $\omega_0$ and at sidebands
$\omega_0+n\omega_{AC}$. Sidebands with $n=\pm 1, \pm 2$ are visible. 
${\cal{K}}^{-+}(\Omega)$ vanishes at zero frequency.
For frequencies larger than $\Omega=\omega_0+2\omega_{AC}$, this noise derivative 
correlator seems to be negligible.
While the formulas for ${\cal{K}}^{-+}(\Omega)$ show a power law divergence, 
here one has to add a regularization procedure because strickly speaking, 
the calculations have been performed in the weak backscattering regime. 
This means that the differential conductance associated with the tunneling current
has to be lower than the conductance quantum (otherwise, one should examine the case of 
the crossover to the strong back scattering regime). The validity condition of our 
results have been previously derived in Eq. (24) of Ref. \onlinecite{crepieux_photoassisted}.  
For our purpose, it just implies that the finite frequency PA noise saturates at 
locations $\Omega=\omega_0+ n\omega_{AC}$.

In the integer quantum Hall case $\nu=1$, no divergences are found for 
${\cal{K}}^{-+}(\Omega)$. Instead, singularities in the derivative 
occur for $\omega_0+n\omega_{AC}$, and the current derivative correlator
seems again to be negligible again beyond $\Omega=\omega_0+2\omega_{AC}$.

\begin{figure}[ht]
	\centering
		\includegraphics[width=18cm]{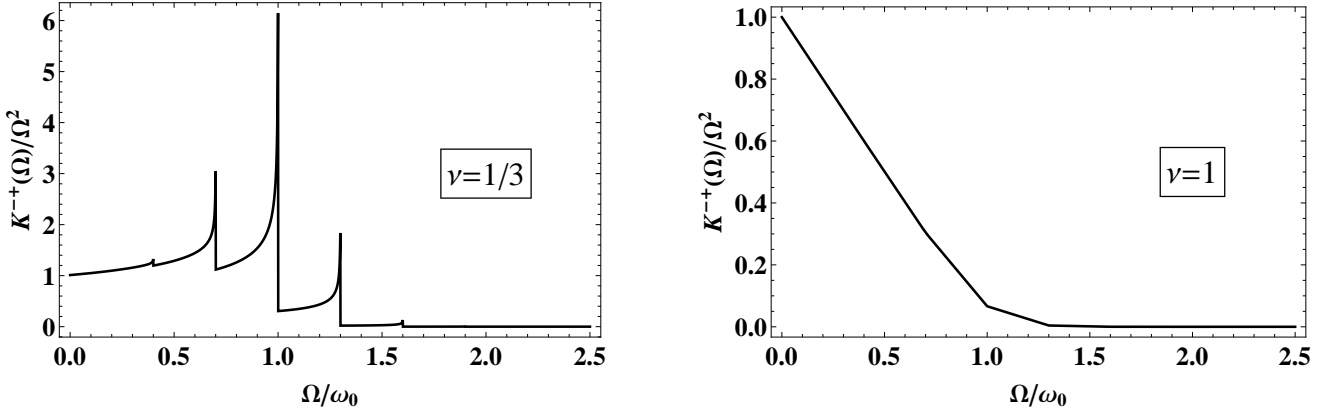}
	\caption{Averaged current correlator
	(same parameters as in Fig. \ref{fig:Bruitderive} except the fact that ${\cal{K}}^{-+}(\Omega)/\Omega^2$ is normalized by $e^* I_B$)}
	\label{fig:Bruit}
\end{figure}

However it is also interesting to plot ${\cal{K}}^{-+}(\Omega)/\Omega^2$; in this way we have access to an ``averaged'' current correlator (noise)
because the term $\Omega^2$ in ${\cal{K}}^{-+}(\Omega)$ is in fact due to derivative operators
acting on the current correlator. This is depicted in Fig. \ref{fig:Bruit}.   
For $\nu=1/3$ we again find divergences (at the same locations as for ${\cal{K}}^{-+}(\Omega)$).
The only noticeable difference with the latter curves is that the averaged noise
does not vanish at zero frequency. If one ignores the side bands, the central peak
reminds us clearly of the finite frequency non symmetrized noise computed recently for 
a QPC in the FQHE.\cite{safi_bena}

For $\nu=1$ the finite frequency noise again
exhibits jumps in its derivative with respect to frequency, but its behavior is linear 
between two successive singularities.
Thus, for this ratio of frequencies $\omega_0/\omega_1>1$, the excess noise characteristics resembles
the finite frequency noise in the absence of an AC drive: the later is (essentially) linear for 
$\Omega<\omega_0$ and vanishes beyon this. Yet the PA noise does not vanish at 
$\Omega=\omega_0$, it shows a singularity in its derivative at its location, together
with singularities at $\Omega=\omega_0+ n\omega_{AC}$ ($n=\pm 1$ is visible). To normalize the curves in Fig. (\ref{fig:Bruitderive}), Fig. (\ref{fig:Bruit}) and in the following section, we use the back scattering current to zero order in the amplitude of $\omega_1$ of the modulation wich corresponds to the pure stationnary regime \cite{crepieux_photoassisted}:

\begin{equation}
	I^{(0)}_B=\frac{e^*\Gamma_0^2}{2\pi^2a^2\Gamma(2\nu)}(\frac{a}{\nu_{F}})^{2\nu}\mathrm{sgn}(\omega_0)\left|\omega_0\right|^{2\nu-1}.
\end{equation}

\subsection{Measured PA noise}

In this section, we display curves for the charge correlator 
at equal times. We consider excess quantities.
By excess, we mean that the charge correlator at zero voltage has 
been subtracted from the charge correlator at $V_0,V_1\neq 0$. 

\begin{figure}[ht]
	\centering
		\includegraphics[width=18cm]{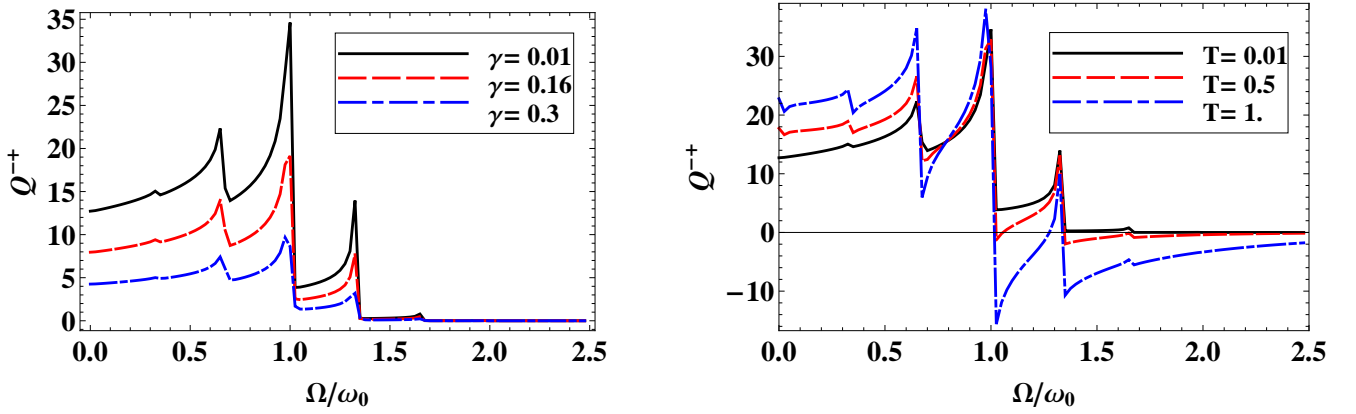}
	\caption{(Color online) PA noise for constant temperature $T=0.01 \omega_0$ (left) and constant dissipation $\gamma=0.01 \omega_0$ (right): 
	$\omega_{AC}=\omega_0/3$, $\omega_1/\omega_{AC}=1$. $Q^{-+}$ is normalized by ${\cal{L}}^2/(\alpha^2 e^* I_B \omega^2_0)$.}
	\label{fig:Twoplot}
\end{figure}

\begin{figure}[ht]
	\centering
		\includegraphics[width=18cm]{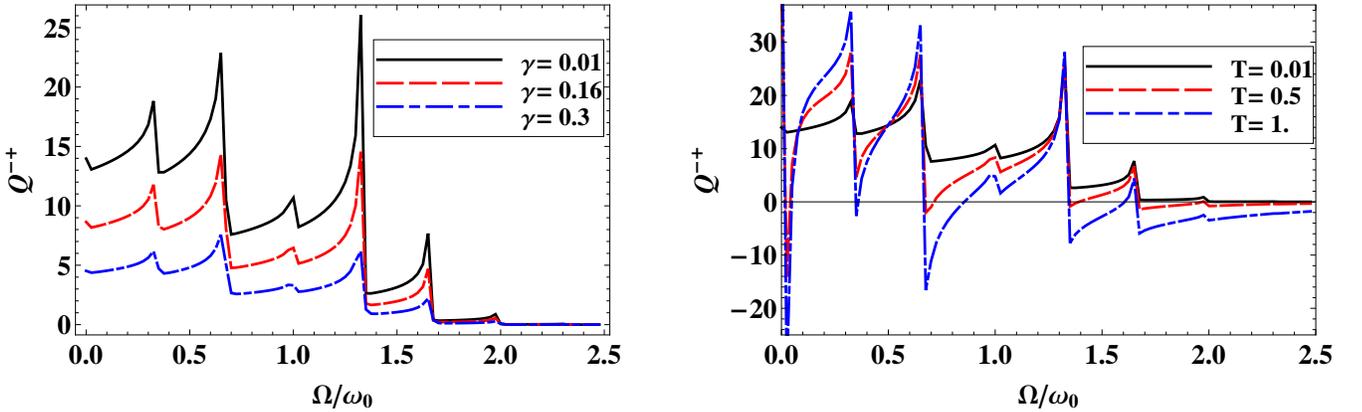}
	\caption{(Color online) PA noise for constant temperature $T=0.01 \omega_0$ (left) and constant dissipation $\gamma=0.01 \omega_0$ (right): 
	$\omega_{AC}=\omega_0/3$,	$\omega_1/\omega_{AC}=2$. The normalization is the same that in Fig. \ref{fig:Twoplot}. }
	\label{fig:Twoplot1}
\end{figure}

\begin{figure}[h]
	\centering
		\includegraphics[width=17cm]{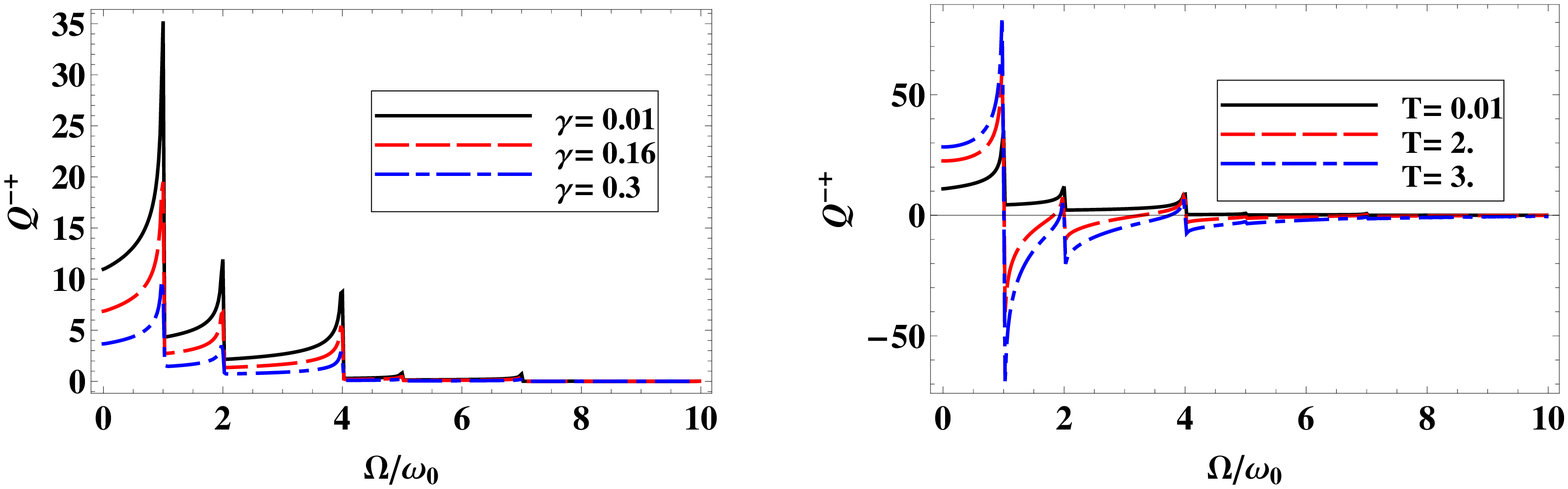}
	\caption{(Color online) PA noise for constant temperature $T=0.01 \omega_0$ (left) and constant dissipation $\gamma=0.01 \omega_0$ (right): 
	$\omega_{AC}=3\, \omega_0$, $\omega_1/\omega_{AC}=1$. The renormalization is the same that in Fig. \ref{fig:Twoplot}.}
	\label{fig:Twoplot2}
\end{figure}

\begin{figure}[h]
	\centering
		\includegraphics[width=18cm]{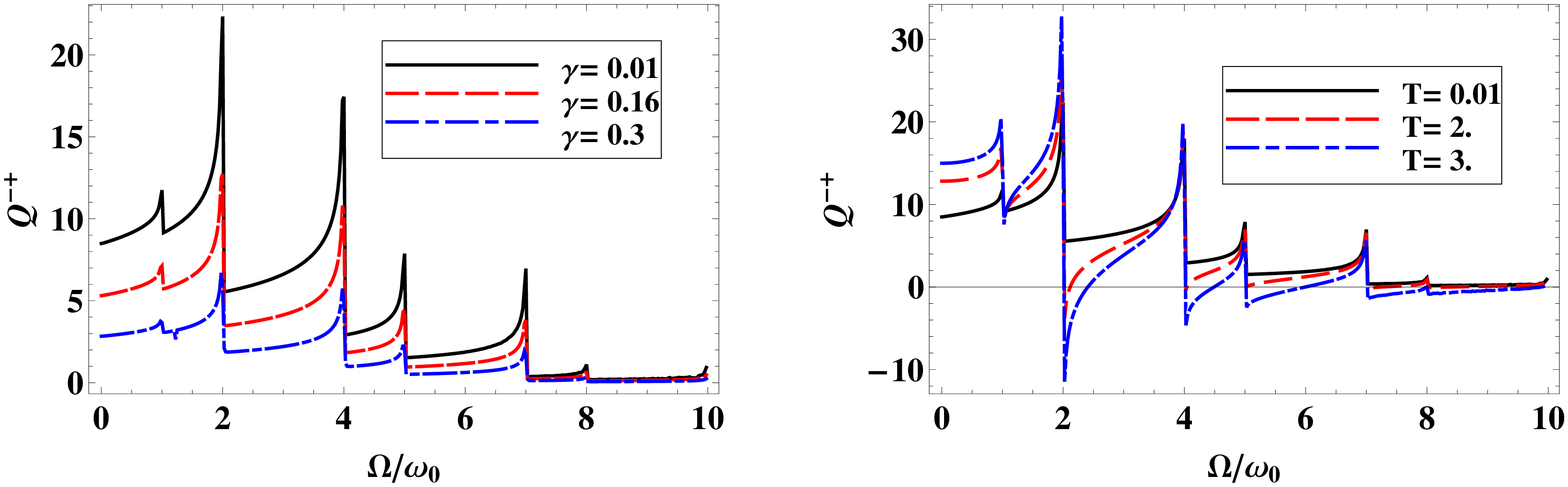}
	\caption{(Color online) PA noise for constant temperature $T=0.01 \omega_0$ (left) and constant dissipation $\gamma=0.01 \omega_0$ (right): 
	$\omega_{AC}=3\, \omega_0$, $\omega_1/\omega_{AC}=2$. The normalization is the same that in Fig. \ref{fig:Twoplot}.}
	\label{fig:Twoplot3}
\end{figure}

In Figs. \ref{fig:Twoplot}, \ref{fig:Twoplot1}, \ref{fig:Twoplot2} and \ref{fig:Twoplot3} we plot these quantities: 

\begin{equation}
\hat{Q}^{-+}(0)=\alpha^2\sum_{\beta_1\beta_2} \int\frac{d\omega}{2\pi}G^{-\beta_2}(\omega)\sigma^{\beta_2\beta_2}_z {\cal{K}}^{\beta_2\beta_1}(\omega)\sigma^{\beta_1\beta_1}_z G^{\beta_1+}(\omega).
\end{equation}

with weak and strong dissipation, low and high (detector) temperature for two different values of the ratio $\omega_1/\omega_{AC}$, which corresponds
to the argument of the Bessel functions in the expression of the charge correlator. In the following curves $Q^{-+}$ is always normalized by ${\cal{L}}^2/(\alpha^2 e^* I_B \omega^2_0)$ and dissipation and temperature are in unit of $\omega_0$. The frequency $\Omega=\omega_0$ corresponds to the positions of the central peak.  

We start with $\omega_0>\omega_{AC}$. 
In order to resolve these peaks, it is necessary that the width 
of the resonance level is smaller than the spacing $\omega_{AC}$ between peaks.  
We observe that by varying $\omega_1/\omega_{AC}$, 
the relative amplitudes of the peaks can be modulated.  

In Fig. \ref{fig:Twoplot1}, the curves correspond to a ratio $\omega_1/\omega_{AC}=2$: we can clearly identify 
the central peak at $\Omega=\omega_0$ but it is smaller than in the case 
$\omega_1/\omega_{AC}=1$ (in Fig. \ref{fig:Twoplot}). 
In this situation we identify very clearly the first and the second satellite peaks,
while the third one ($n=\pm 3$) is visible but with a lesser intensity. The relative amplitude of the central peak and its satellite is tied to the oscillatory behavior of the Bessel function.

When $\omega_1/\omega_{AC}=1$, the $0$th order Bessel function, which corresponds 
to the central peak has a large amplitude($\approx 0.6$). The $1$st order Bessel function which corresponds to the first satellite peak, has a smaller amplitude ($\approx 0.2$). 
The third Bessel function which corresponds to the second satellite peak is almost zero. 
On the other hand for $\omega_1/\omega_{AC}=2$, the $0$th and the $3$rd order 
Bessel function are small compared to its $1$st and $2$nd order counterparts, 
thus the central peak is smaller than the satellites. 

Next, we choose $\omega_0<\omega_{AC}$ in Fig. \ref{fig:Twoplot2} and \ref{fig:Twoplot3}. The finite frequency spectrum 
of charge fluctuations does not seem to display any longer a central peak with 
equally spaced satellites.   

In Fig. \ref{fig:Twoplot2}, the curves correspond to a ratio $\omega_1/\omega_{AC}=1$ and in Fig. \ref{fig:Twoplot3} the curves correspond to a ratio $\omega_1/\omega_{AC}=2$.
In Fig. \ref{fig:Twoplot2}, the curves have a central peak at frequency $\omega_0$, a secondary one at $\omega_0+\omega_{AC}$ and a third one at $\omega_0+2\omega_{AC}$. 
However there appear peaks at frequencies $-\omega_0+\omega_{AC}$ and $-\omega_0+2\omega_{AC}$: this corresponds to the satellites peaks 
of the negative frequency $-\omega_0$. We can explain this phenomena as the overlapping of two combs, 
centered at $\pm \omega_0$. In Fig. \ref{fig:Twoplot3}, the curves exhibit the same phenomena but the peaks 
have different relative amplitude, which can again be explained from the argument of 
the Bessel functions. In these differents plots, we can see on the one hand the effect of dissipation wich reduces the noise and smoothes the peaks. On the other hand we see the effect of temperature; the measured noise become negative at higher temperature (as in the case $\nu=1$ and $DC$ applied voltage \cite{zazunov}) because of $S_{+}-S_{-}<0$ wich is larger than $S_{+}$ and because of the large population of LC oscillator states. 
  
\section{Conclusion}

The central point of this paper has been the presentation of a measurement 
scheme for detecting finite frequency photo-assisted noise of a mesoscopic 
conductor on which both a DC and an AC bias is imposed. This scheme uses a 
dissipative resonant circuit which is inductively compled to the mesoscopic 
circuit, in the same manner as some of the author's previous work\cite{zazunov}, 
which we reviewed at the begining of the paper. The major hurdle in analyzing PA shot 
noise lies in the lack of time translational invariance which results from the 
presence of the AC drive. We have shown that by considering the average of the 
charge correlator of the LC over the period of the AC drive, time translational 
invariance can be restored, and an extension of our previous detection scheme
can be envisioned. 

We illustrated our detection scheme by applying it it to a concrete situation where PA noise
features are most visible.  
We therefore considered the PA noise generated from a 
point contact in the weak-backscattering regime, placed in the regime of the FQHE. 
While the symmetrized PA noise at zero frequency was previously derived 
by some of us, no derivation of its full Keldysh components at finite frequency
was available to this date. The PA noise contains singularities at frequencies
corresponding to the bias voltage, with satellite singularities separated by the 
AC drive frequency. These sharp features in the noise are the main motivation 
for the application of our detection scheme. 
After deriving in this situation the current derivative correlator, we were able to 
compute explicitly the response of the LC circuit via the charge correlator, and to 
display the results for a variety of parameters. 

Coupling of the detector to an electromagnetic environment, here modelled by an ohmic bath of oscillators, smoothens the anomalies of the detected signal. The damping parameter ought to be smaller than either the DC frequency or the AC drive frequency in order that the desired effects are observed. This observation is crucial for experiments, and broadens the scope of the results since the electromagnetic environment may also model other backaction effects on the detector.   

The second important and non trivial effect is that the measured noise become negative if we increase the temperature of the detector. Remember that we are considering excess measurements; negative noise thus means that the noise for non-zero DC and AC voltages is smaller that the noise for zero voltage.

Given the fact that the AC modulation gives rise to satellite peaks at $\pm \omega_0+n\omega_{AC}$, we distinguished two limits: $\omega_0 > \omega_{AC}$ where the central peak at the DC voltage
is surrounded by its satellites, and $\omega_0 < \omega_{AC}$ where the satellites of the negative
DC voltage frequency can lie in the positive frequency domain of the charge correlator. 
Both situations can be realized in practice. This brings us to the question about optimizing 
the detection of the location of the central peak and its satellites. Upon varying the ration 
between the AC drive amplitude and the AC frequency, we have shown that one can modify the 
respective amplitude of such peaks. This constitutes an additional knob for detection. 

The present results constitute a step in the direction of fundamental aspects of mesoscopic 
physics of detection in the time domain. This is an area of growing importance in mesoscopic 
physics when conventional detection machinery has to be abandoned, and novel detection schemes
for high frequencies adapted to the type of experiments 
on wishes to perform. Granted, from the experimental point of view, the LC circuit setup 
which we have presented here may appear a bit naive. In the long run, indeed one should attempt
to describe more precisely the connection between the output signal of the mesoscopic 
circuit and the the transmission lines which are connected to it. This will be the topic 
of further investigations.
    
\acknowledgments 
We thank Alex Zazunov, Per Delsing and Tim Duty for useful discussions. T.M. and T.J. acknowledge support of an ANR grant 
``Molspintronics'' from the French ministry of research. T.M., T.J., E.P. and G.F. acklowledge support from a Gallileo 
project of the ``Partenariat Hubert Curien''.  E.P. thanks CPT for its hospitality. T.M. thanks University of Catania
for its hospitality. 

\appendix

\section{Keldysh noise correlator calculation}

From Eq. (\ref{variable_change}) we use a standard trigonometric identity in order to factorize
the noise into contributions with $\tau$ and $\tau'$:
\begin{eqnarray}
S^{\beta\beta'}(\Omega_1,\Omega_2)&=&2\frac{(e^*)^2\Gamma_0^2}{2\pi^2a^2}\sum_{n=-\infty}^{+\infty}
\sum_{m=-\infty}^{+\infty}J_n\left(\frac{e^*V_1}{\omega_{AC}}\right)
J_m\left(\frac{e^*V_1}{\omega_{AC}}\right)\nonumber\\
&~~&~~\left[I_{1}(\Omega_1+\Omega_2,\omega)
I_{2}^{\beta\beta'}(\Omega_1-\Omega_2,\omega_0,\omega)
-
I_{3}(\Omega_1+\Omega_2,\omega)
I_{4}^{\beta\beta'}(\Omega_1-\Omega_2,\omega_0,\omega)
\right]
~,\end{eqnarray}
with:
\begin{eqnarray}
I_{1}(\Omega_1+\Omega_2,\omega)&=&\int_{-\infty}^{+\infty}  d\tau'e^{i(\Omega_1+\Omega_2)\tau'/2}\mathrm{cos}\left(\frac{n-m}{2}\omega_{AC}\tau'\right)
\nonumber\\
I_{2}^{\beta\beta'}(\Omega_1-\Omega_2,\omega_0,\omega)
&=&
\int_{-\infty}^{+\infty} d\tau e^{i(\Omega_1-\Omega_2)\tau/2}
e^{2\nu G^{\beta\beta'}(\tau)}\mathrm{cos}\left(\left(\omega_0+\frac{n+m}{2}\omega_{AC}\right)\tau\right)
\nonumber\\
I_{3}(\Omega_1+\Omega_2,\omega)
&=&
\int_{-\infty}^{+\infty}  d\tau'e^{i(\Omega_1+\Omega_2)\tau'/2}\mathrm{sin}\left(\frac{n-m}{2}\omega_{AC}\tau'\right)
\nonumber\\
I_{4}^{\beta\beta'}(\Omega_1-\Omega_2,\omega_0,\omega)
&=&
\int_{-\infty}^{+\infty}  d\tau e^{i(\Omega_1-\Omega_2)\tau/2}
e^{2\nu  G^{\beta\beta'}(\tau)}\mathrm{sin}\left(\left(\omega_0+\frac{n+m}{2}\omega_{AC}\right)\tau\right)
~,\end{eqnarray}
with the elements of the Keldysh Green's function for the chiral field:
\begin{eqnarray}
G^{\beta \beta}(\tau)&=&-ln\left(1+\beta i\frac{\nu_{F}\left|\tau\right|}{a}\right)\\
G^{\beta-\beta}(\tau)&=&-ln\left(1-\beta i\frac{\nu_{F}\tau}{a}\right)~.
\end{eqnarray}

$I_1$ and $I_3$ are expressed in terms of delta functions:
\begin{equation}
I_{1}=\frac{1}{2}\left(\delta(\Omega_1+\Omega_2+(n-m)\omega_{AC})+\delta(\Omega_1+\Omega_2-(n-m)\omega_{AC})\right)
~.\end{equation}
\begin{equation}
I_{3}=\frac{1}{2i}\left(\delta(\Omega_1+\Omega_2+(n-m)\omega_{AC})-\delta(\Omega_1+\Omega_2-(n-m)\omega_{AC})\right)
~.\end{equation}

Integrals $I_2^{\beta\beta'}$ and $I_4^{\beta\beta'}$ depend explicitly on the Keldysh 
indices $\beta$ and $\beta'$. Here, we
need two tabulated integrals:
\begin{eqnarray}
\int_{-\infty}^{+\infty}\frac{\sin(\omega_0\tau)d\tau}
{\left(\frac{a}{v_F}-i\eta\tau\right)^{\mu}}
&\approx& i\pi\eta\mathrm{sgn}(\omega_0)\frac{|\omega_0|^{\mu-1}}{{\bf
\Gamma}(\mu)}\\
\int_{-\infty}^{+\infty}\frac{\cos(\omega_0\tau)d\tau}
{\left(\frac{a}{v_F}-i\eta\tau\right)^{\mu}}
&\approx& \pi\frac{|\omega_0|^{\mu-1}}{{\bf
\Gamma}(\mu)}
~.\end{eqnarray}
The results for $I_{2}^{\beta\beta'}$ and $I_{4}^{\beta\beta'}$ are:

\begin{align}
I_{2}^{\beta-\beta}
&=\frac{\pi}{2\Gamma(2\nu)}(\frac{a}{\nu_{F}})^{2\nu}\left[\left(1-\beta \mathrm{sgn}\left(\frac{\Omega_1-\Omega_2}{2}-\omega_0-\frac{n+m}{2}\omega_{AC}\right)\right)\left|\frac{\Omega_1-\Omega_2}{2}-\omega_0-\frac{n+m}{2}\omega_{AC}\right|^{2\nu-1}\right.\notag\\
&+\left.\left(1-\beta\mathrm{sgn}\left(\frac{\Omega_1-\Omega_2}{2}+\omega_0+\frac{n+m}{2}\omega_{AC}\right)\right)\left|\frac{\Omega_1-\Omega_2}{2}+\omega_0+\frac{n+m}{2}\omega_{AC}\right|^{2\nu-1}\right]
~.\end{align}

\begin{equation}
I_{2}^{\beta\beta}=\frac{1}{2}(\frac{a}{\nu_F})^{2\nu}\frac{\pi}{\Gamma(2\nu)}\frac{e^{-\beta i\pi\nu}}{\mathrm{cos}(\pi\nu)}\left(\left|\frac{\Omega_1-\Omega_2}{2}-\omega_0-\frac{n+m}{2}\omega_{AC}\right|^{2\nu-1}+\left|\frac{\Omega_1-\Omega_2}{2}+\omega_0+\frac{n+m}{2}\omega_{AC}\right|^{2\nu-1}\right)
~.\end{equation}

\begin{align}
I_{4}^{\beta-\beta}
&=\frac{1}{2}(\frac{a}{\nu_{F}})^{2\nu}\frac{i\pi}{\Gamma(2\nu)}\left[\left(1+\beta\mathrm{sgn}(\omega_0+\frac{n+m}{2}\omega_{AC}-\frac{\Omega_1-\Omega_2}{2})\right)\left|\omega_0+\frac{n+m}{2}\omega_{AC}-\frac{\Omega_1-\Omega_2}{2}\right|^{2\nu-1}\right.\notag\\
&-\left.\left(1+\beta\mathrm{sgn}(\omega_0+\frac{n+m}{2}\omega_{AC}+\frac{\Omega_1-\Omega_2}{2})\right)\left|\omega_0+\frac{n+m}{2}\omega_{AC}+\frac{\Omega_1-\Omega_2}{2}\right|^{2\nu-1}\right]
~.\end{align}

\begin{align}
I_{4}^{\beta\beta}&=\frac{i}{2}(\frac{a}{\nu_{F}})^{2\nu}\frac{\pi e^{-\beta i\pi\nu}}{\Gamma(2\nu)\mathrm{cos}(\pi\nu)}\left|\omega_0+\frac{n+m}{2}\omega_{AC}-\frac{\Omega_1-\Omega_2}{2}\right|^{2\nu-1}\notag\\
&-\frac{i}{2}(\frac{a}{\nu_{F}})^{2\nu}\frac{\pi e^{-i\pi\nu}}{\Gamma(2\nu)\mathrm{cos}(\pi\nu)}\left|\omega_0+\frac{n+m}{2}\omega_{AC}+\frac{\Omega_1-\Omega_2}{2}\right|^{2\nu-1}
~.\end{align}

\section{Time average current derivative correlators in the rotated Keldysh basis}\label{rakbasis}

Here, for completeness we compute the 
components ${\cal{K}}^R$, ${\cal{K}}^{A}$ and ${\cal{K}}^K$
in the rotated Kelsysh basis.
We recall that if the time ordered Keldysh components of any correlator (such as the LC Greens function) read:
\begin{equation}
\hat{G}=\begin{pmatrix}
G^{++}&	G^{+-}\\
G^{-+}&	G^{--}
\end{pmatrix}~,
\end{equation}
then the rotated Kelysh matrix is defined as:
\begin{equation}
\tilde{G}=L\tau_{z}\hat{G}L^{-1}=\begin{pmatrix}
G^{R}&	G^{K}\\
0&	G^{A}
\end{pmatrix}~.
\end{equation}
where $L$ is the unitary transformation:
\begin{equation}
\frac{1}{\sqrt{2}}\begin{pmatrix}
1& 	-1\\
1&	1
\end{pmatrix}~.
\end{equation}
We obtain from the expressions of the previous section:

\begin{align}
&\tilde{{\cal K}}^{R/A}(\omega)=\frac{1}{\tau}\int^{\tau}_{0}dT\tilde{k}^{R}(T,\omega)=\frac{(e^*)^2\Gamma_0^2}{4\pi^2a^2}\sum^{+\infty}_{n=-\infty}J^2_n\left(\frac{e^*V_1}{\omega_{AC}}\right)\frac{1}{\Gamma(2\nu)}(\frac{a}{\nu_{F}})^{2\nu}\omega^2\times\notag\\
&\left[\left(-i\mathrm{tan}(\pi\nu)\pm\mathrm{sgn}\left(\omega+\omega_0+n\omega_{AC}\right)\right)\left|\omega+\omega_0+n\omega_{AC}\right|^{2\nu-1}+\left(-i\mathrm{tan}(\pi\nu)\pm\mathrm{sgn}\left(\omega-\omega_0-n\omega_{AC}\right)\right)\left|\omega-\omega_0-n\omega_{AC}\right|^{2\nu-1}\right]
~.\end{align}

\begin{align}
&\tilde{{\cal K}}^{K}(\omega)=\frac{1}{\tau}\int^{\tau}_{0}dT\tilde{k}^{K}(T,\omega)=\frac{(e^*)^2\Gamma_0^2}{4\pi^2a^2}\sum^{+\infty}_{n=-\infty}J^2_n\left(\frac{e^*V_1}{\omega_{AC}}\right)\frac{1}{\Gamma(2\nu)}(\frac{a}{\nu_{F}})^{2\nu}\omega^2\times\notag\\
&2\left[\left|\omega+\omega_0+n\omega_{AC}\right|^{2\nu-1}+\left|\omega-\omega_0-n\omega_{AC}\right|^{2\nu-1}\right]~.
\end{align}

Turning now to the charge correlator at equal time, its matrix expression yields in the 
time ordered basis:
\begin{equation}
\hat{Q}(0)=\alpha^2 \int\frac{d\omega}{2\pi}\hat{G}(\omega)\sigma_z\hat{\cal K}(\omega)\sigma_z\hat{G}(\omega)~.
\end{equation}
or in the rotated basis:
\begin{align}
&\tau_zL^{-1}\tilde{Q}L=\hat{Q} 
&\tilde{Q}=\alpha^2 \int\frac{d\omega}{2\pi}\tilde{G}(\omega)\tilde{{\cal{K}}}(\omega)\tilde{G}(\omega)~.
\end{align}
This allow to obtain the Keldysh rotated elements of the charge correlator:
\begin{equation}
\tilde{Q}^{R/A}=\alpha^2\int\frac{d\omega}{2\pi}\tilde{G}^{R/A}(\omega)\tilde{{\cal{K}}}^{R/A}(\omega)\tilde{G}^{R/A}(\omega)~.
\end{equation}

\begin{equation}
\tilde{Q}^K=\alpha^2\int\frac{d\omega}{2\pi}\left[\tilde{G}^R(\omega)\tilde{{\cal{K}}}^R(\omega)\tilde{G}^K(\omega)+\tilde{G}^R(\omega)\tilde{{\cal{K}}}^K(\omega)\tilde{G}^A(\omega)+\tilde{G}^K(\omega)\tilde{{\cal{K}}}^A(\omega)\tilde{G}^A(\omega)\right]~.
\end{equation}

\end{document}